\documentclass[preprint,aps,floatfix,nofootinbib,a4paper]{revtex4-1}

\pdfoutput=1

\usepackage{amsmath, amssymb, amsfonts, amsthm, latexsym, epsfig, mathrsfs, xcolor, bbm, slashed}

\usepackage[all]{xy}

\usepackage[inline]{enumitem}

\usepackage{multirow}

\usepackage{setspace}
\usepackage[marginal, multiple]{footmisc}

\usepackage[T1]{fontenc}
\usepackage[utf8]{inputenc}
\usepackage{lmodern}

\usepackage[colorlinks, allcolors=blue!70!black, linktocpage]{hyperref}

\numberwithin{equation}{section}

\usepackage{cleveref}

\usepackage{microtype}

\usepackage{floatrow}

\let\OLDtableofcontents\tableofcontents
\renewcommand\tableofcontents[1]{%
    {\baselineskip 0.5ex %
	\OLDtableofcontents{#1}}%
}

\let\OLDthebibliography\thebibliography
\renewcommand\thebibliography[1]{%
	\setstretch{1.079} 
	\OLDthebibliography{#1}%
	\small %
	\setlength{\itemsep}{0.2\baselineskip} 
}

\let\OLDfootnote\footnote
\renewcommand\footnote[1]{%
	\setlength{\footnotesep}{0.75\baselineskip}%
	{\footnotesize \OLDfootnote{#1}}%
}

\setlist[enumerate]{noitemsep, label=(\arabic*), ref=(\arabic*)}

\newlist{condlist}{enumerate}{2}
\setlist[condlist,1]{noitemsep, topsep=0pt, label=(\arabic*), ref=(\arabic*)}
\setlist[condlist,2]{noitemsep, label=(\alph*), ref=(\arabic{condlisti}.\alph*)}
\crefname{condlisti}{condition}{conditions}
\crefname{condlistii}{condition}{conditions}

\newlist{propertylist}{enumerate}{1}
\setlist[propertylist,1]{noitemsep, topsep=0pt, label=(\arabic*), ref=(\arabic*)}
\crefname{propertylisti}{Property}{Properties}

\renewcommand\thesection{\arabic{section}}
\renewcommand\thesubsection{\arabic{subsection}}

\makeatletter
\def\p@subsection{\thesection.}
\def\p@subsubsection{\thesection.\thesubsection.}
\makeatother 

\theoremstyle{plain}
\newtheorem{thm}{Theorem}
\newtheorem{lemma}{Lemma}[section]
\newtheorem{prop}{Proposition}[section]

\theoremstyle{definition}

\theoremstyle{remark}
\newtheorem{remark}{Remark}[section]

\crefname{section}{\S}{\S}
\crefname{figure}{Fig.}{Figs.}
\crefname{table}{Table}{Tables}

\crefname{definition}{Def.}{Defs.}
\crefname{prop}{prop.}{props.}

\crefname{ass}{Assumptions}{Assumptions}
\crefname{property}{Properties}{Properties}

\newcommand{\be}{\begin{equation}\begin{aligned}}
\newcommand{\ee}{\end{aligned}\end{equation}}

\newcommand{\lb}{\left}
\newcommand{\rb}{\right}

\newcommand{\lra}{\leftrightarrow}

\newcommand{\ms}{\mathscr}
\newcommand{\mf}{\mathfrak}
\newcommand{\bb}{\mathbb}

\newcommand{\eqsp}{\, ,\quad} 

\newcommand{\hr}{\begin{center}* * *\end{center}}

\newcommand{\Lie}{\pounds} 
\newcommand{\defn}{\mathrel{\mathop:}=} 

\newcommand{\inter}{\cap} 

\let\oldsetminus\setminus
\renewcommand{\setminus}{\!\oldsetminus\!} 

\let\oldint\int
\renewcommand{\int}{\oldint\limits}

\let\oldlim\lim
\renewcommand{\lim}{\oldlim\limits}

\renewcommand{\bar}{\overline}

\newcommand{\scri}{\ms I}

\newcommand{\hateq}{\mathrel{\mathop {\widehat=} }} 

\newcommand{\nfrac}[2]{{{}^#1\!\!/\!_#2}}

\newcommand{\half}{\nfrac{1}{2}}

\newcommand{\pb}[1]{\underleftarrow{#1}} 

\renewcommand{\Re}{{\rm Re\,}}
\renewcommand{\Im}{{\rm Im\,}}
\newcommand{\abs}[1]{\lb\vert\, #1 \,\rb\vert}		

\let\thorn\relax
\DeclareMathOperator{\thorn}{\text{\rm \th}}
\let\eth\relax
\DeclareMathOperator{\eth}{\text{\rm \dh}}

\newcommand{\wt}{\circeq}

\renewcommand{\o}{o}
\renewcommand{\i}{\iota}

\newcommand{\bms}{\mf{bms}}

\begin{document}

\setstretch{1.2}


\title{A novel supersymmetric extension of BMS symmetries at null infinity}

\author{Kartik Prabhu}
\email{kartikprabhu@ucsb.edu}
\affiliation{Department of Physics, University of California, Santa Barbara, CA 93106, USA}

\begin{abstract}
We show that we can combine the (complex, self-dual) BMS vector fields with the recently defined BMS twistors to obtain a new supersymmetric extension of the BMS symmetries at null infinity. We compare our construction to other supersymmetric extensions of the BMS algebra proposed in the context of supergravity. Unlike the standard constructions the anticommutator in our superalgebra generates all the BMS vector fields including the Lorentz transformations. We also show that there exists a projection from our BMS Lie superalgebra to the global subalgebra of the Neveu-Schwarz supersymmetries on a \(2\)-sphere, which are commonly considered in string theory and \(2\)-dimensional conformal field theory.
\end{abstract}

\maketitle
\tableofcontents

\section{Introduction}
\label{sec:intro}

Following up on the recent construction of Bondi-Metzner-Sachs (BMS) twistors at null infinity \cite{bms-twistors}, we describe how these twistors and BMS symmetries can be unified into a supersymmetric algebra, that is a Lie superalgebra.

For asymptotically flat spacetimes describing isolated systems in general relativity, it is well-known that at the asymptotic boundary, \emph{null infinity} denoted by \(\scri\), one obtains an infinite-dimensional asymptotic symmetry group --- the BMS group --- along with the corresponding charges and fluxes due to gravitational radiation \cite{BBM, Sachs1, Sachs2, Penrose, Geroch-asymp,GW, AS-symp, WZ}, see also \cite{GPS} for a recent exposition. We review the properties of the BMS vector fields in \cref{sec:bms-vectors}.

The BMS twistors in \cite{bms-twistors} were motivated by obtaining a twistorial description of the BMS Lie algebra. As is well-known it is not possible to impose even the tangential components of the twistor equation on general spacetimes at \(\scri\), unless the News (i.e. any gravitational radiation) vanishes. The usual strategy in the twistor literature is to select a fixed cross-section of \(\scri\) and only impose those components of the twistor equation which are tangent to this cross-section; this defines the \emph{\(2\)-surface twistors} on a cross-section of \(\scri\). These \(2\)-surface twistors can be used to generate a Poincar\'e algebra at the chosen cross-section \cite{Penrose-charges, DS, Shaw}. The alternative strategy used in \cite{bms-twistors} is to instead impose those components of the full twistor equation which are \emph{both} tangent and universal on \(\scri\). The infinite-dimensional space of solutions to these equations can be used to generate (complex) BMS vector fields; this construction is recalled in \cref{sec:bms-twistor}.

Most of the structures to elevate this construction to a Lie superalgebra have already been defined --- the BMS vector fields form a Lie algebra which will be the even part of our Lie superalgebra; the symmetric map from the BMS twistors to a BMS vector field defined in \cite{bms-twistors} defines a bracket on the odd part of the Lie superalgebra. The only missing part is a bracket between the BMS vector fields and the BMS twistors, i.e., an action of BMS symmetries on the space of BMS twistors; we will define this action in \cref{sec:PRLie} using a Lie derivative on spinor fields defined by Penrose and Rindler \cite{PR2}. In \cref{sec:superalgebra} we prove that these brackets satisfy the Jacobi identities and hence define a Lie superalgebra \(\mf K\). We also compare and contrast this Lie superalgebra with other constructions of supersymmetries at null infinity. We also describe a curious aspect of the Lie superalgebra \(\mf K\): the projection of \(\mf K\) to the space of null generators of \(\scri\) reproduces the (global) Neveu-Schwarz supersymmetric algebra (see \cref{sec:NS}). We conclude with a short discussion of some interesting new avenues suggested by this work in \cref{sec:disc}.

\section*{Notation and conventions}

Abstract indices \(a,b,\ldots\) will be used for tensors in spacetime while \(A,B,\ldots\) and \(A',B',\ldots\) will be used for abstract spinor indices using the conventions in \cite{PR1}. We work exclusively in the conformally completed spacetime, the \emph{unphysical} spacetime \(M\) with a Lorentzian metric \(g_{ab}\). We use the mostly negative signature \((+,-,-,-)\) for the Lorentzian \(4\)-dimensional metric tensor \(g_{ab}\) on spacetime and denote the corresponding (antisymmetric) metrics on the spinor spaces by \(\epsilon_{AB}\) and \(\epsilon_{A'B'}\), see \cite{PR1}. We will use the sign conventions of \cite{PR1} for the Riemann tensor (this is the opposite sign compared to the convention in Wald \cite{Wald-book}); so if \(v_a\) is a \(1\)-form we have
\be\label{eq:Riem-defn}
    \nabla_a \nabla_b v_c - \nabla_b \nabla_a v_c = - R_{abc}{}^d v_d
\ee
Since the Riemann tensor is antisymmetric in the last two indices we also have
\be\label{eq:Riem-spinor}
    R_{abcd} = R_{abCD} \epsilon_{C'D'} + R_{abC'D'} \epsilon_{CD}
\ee

We also use \(\hateq\) to denote equality at null infinity.

\section{Null infinity, BMS symmetries and BMS twistors}
\label{sec:bms}

We will use the definition of asymptotic flatness given by Penrose's conformal completion (see \cite{Geroch-asymp,Wald-book}), and denote null infinity as \(\scri \cong \bb R \times \bb S^2\). Let \(\Omega\) be the conformal factor used to obtain the conformal-completion of the physical spacetime, then it can be shown that \(\nabla^a\Omega\) is the null generator of \(\scri\), and that one can, without loss of generality, choose \(\Omega\) so that the Bondi condition \(\nabla_a \nabla_b \Omega \hateq 0\) is satisfied at \(\scri\). For intermediate computations, we will use the Geroch-Held-Penrose (GHP) formalism at null infinity \cite{GHP,PR1,PR2,KP-GR-match}. The GHP weight of any quantity \(\eta\) will be denoted by \(\eta \wt (p,q)\), and its spin will be \(s = (p-q)/2\). For this it will be convenient to make a choice of a null tetrad and spinor basis at \(\scri\) which determines a \emph{Bondi system}, see \cite{PR2} for details.

We pick a vector field \(n^a\) and a spinor \(\i^A\) at null infinity so that
\be\label{eq:normal-defn}
    A n^a \hateq - \nabla^a\Omega \eqsp n^a \hateq \i^A \i^{A'}
\ee
for some function \(A\) with GHP weights \(A \wt (1,1)\). Next, we pick a foliation of \(\scri\) so that the cross-sections are parallely-transported along \(n^a\). This foliation determines a unique null vector field \(l^a\) at \(\scri\) so that \(l_a \hateq g_{ab}l^b\) is the conormal to the cross-sections and \(n_a l^a \hateq 1\). Finally, we pick a complex null basis \(m^a\) and \(\bar m^a\) which is tangent to the cross-sections of this foliation and \(m_a \bar m^a \hateq -1\). In this basis,
\be
    g_{ab} \hateq 2 n_{(a} l_{b)} - 2 m_{(a} \bar m_{b)} \eqsp q_{ab} \hateq - 2 m_{(a} \bar m_{b)} \eqsp \varepsilon_{ab} \hateq - 2 i m_{[a} \bar m_{b]}
\ee
where \(q_{ab}\) is the pullback of \(g_{ab}\) to \(\scri\), and is a (negative definite) Riemannian metric on the cross-sections of \(\scri\), and \(\varepsilon_{ab}\) is the area-element. We can also define another spinor \(\o^A\) so that \((\o^A, \i^A)\) and their complex conjugates \((\o^{A'}, \i^{A'})\) are associated with the tetrads in the usual way (see \cite{PR2} for details) and are normalized so that
\be
    \o_A \i^A \hateq \o_{A'} \i^{A'} \hateq 1
\ee
and all other contractions vanishing. In this choice of basis, the GHP spin coefficients at \(\scri\) satisfy
\be\label{eq:spin-scri}
    \kappa' \hateq \sigma' \hateq \tau' \hateq \rho' \hateq \tau \hateq \Im \rho \hateq 0
\ee
while the spin coefficients \(\kappa, \sigma, \Re \rho\) are arbitrary. The function \(A\) appearing in \cref{eq:normal-defn} satisfies (see Eq.~9.8.26 of \cite{PR2})
\be\label{eq:A-relations}
    \thorn'A \hateq \eth A \hateq 0 \,.
\ee
Note that the spin coefficients \(\kappa\) and \(\Re \rho\) can also be set to zero, by appropriate choices of the conformal factor and tetrad away from \(\scri\), but we will not need to do so. The only non-trivial spin coefficient at \(\scri\) is \(\sigma\) which encodes the gravitational radiation through the News tensor, which is represented by a complex function \(N\) with
\be\label{eq:News-defn}
    \bar N \defn \thorn' \sigma
\ee

\hr

In the following we summarize the universal structure induced on null infinity as the conformal boundary of an asymptotically-flat spacetime; see \cite{Geroch-asymp,AS-symp}. Note we will retain the function \(A \wt (1,1)\) introduced in \cref{eq:normal-defn} to keep track of the GHP weights in our choice of tetrad basis; if one is concerned only with tensorial expressions then \(A\) can be set to \(1\).

Let us recall the ``first-order'' structure of \(\scri\) consists of a vector field \(A n^a\) and a degenerate Riemannian metric \(q_{ab}\), such that \((An^a) q_{ab} \hateq 0\). This structure is universal, in the sense that \(n^a\) and \(q_{ab}\) are intrinsically defined on the manifold \(\scri\), and are common to all asymptotically flat spacetimes. Different asymptotically flat spacetimes are instead distinguished by the ``second-order'' structure encoded in equivalence classes of derivative operators on \(\scri\); we recall the essential aspects below and refer to \cite{Geroch-asymp,AS-symp} for details.

Let \(v_a\) be a \(1\)-form on \(\scri\) and let \(\tilde v_a\) be \emph{any} extension of \(v_a\) into the spacetime \(M\), i.e. \(\tilde v_a\) is a \(1\)-form in \(M\) such that \(v_a = \pb{\tilde v_a}\), where \(\pb{}\) denotes the pullback to \(\scri\). Then, a derivative operator \(D_a\) on \(\scri\) is defined as (see pp.~46 of \cite{Geroch-asymp})
\be
    D_a v_b \defn \pb{\nabla_a \tilde v_b}
\ee
Note that \(D_a\) is well-defined since it is independent of the choice of extension \(\tilde v_a\) of \(v_a\) into the spacetime \(M\), i.e. replacing \(\tilde v_a\) with \(\tilde v_a + \nu A n_a + \Omega \lambda_a\) does not affect \(D_a v_b\) on \(\scri\) \cite{Geroch-asymp}. Intrinsically on \(\scri\), this derivative operator satisfies
\be
    D_a (An^b) \hateq 0 \eqsp D_a q_{bc} \hateq 0
\ee

Two derivative operators \(\hat D_a\) and \(D_a\) are equivalent (they represent different conformal completions of the \emph{same} physical spacetime) if \cite{AS-symp}
\be
    (\hat D_a - D_a)v_b \hateq f q_{ab} (An^c) v_c \hateq - (\hat \rho - \rho) q_{ab} (An^c) v_c 
\ee
for some function \(f\) and \emph{all} \(v_b\) on \(\scri\). In our tetrad basis this function is given by the difference of the spin coefficient \(\rho\) as indicated above. Let us denote by \(\{D\}_a\) the equivalence class of the derivative operator \(D_a\) under the above equivalence relation.

The difference of equivalence classes of derivatives is given by a tensor \(\gamma_{ab}\)
\be
    \lb( \{\hat D\}_a - \{D\}_a \rb) v_b \hateq \gamma_{ab} (A n^c) v_c  \eqsp \gamma_{ab} (An^b) \hateq 0 \eqsp q^{ab} \gamma_{ab} \hateq 0
\ee
for \emph{all} \(v_b\). In our tetrad basis this is
\be
    \gamma_{ab} \hateq (\hat\sigma - \sigma) \bar m_a \bar m_b + c.c.
\ee
where \(c.c.\) denotes the complex conjugate of the previous expression. The shear spin coefficient \(\sigma\) encodes the different equivalence classes of derivatives and thus the radiative degrees of freedom at \(\scri\) \cite{AS-symp}.

We can extend these considerations to spinor fields defined on \(\scri\). Note that since \(n^a = \i^A \i^{A'}\), we can consider \(\i^A\) and its conjugate as part of the universal structure. We can easily extend the derivative operator \(D_a\) to act on spinor fields on \(\scri\) as follows. Let \(\mu^A\) be any spinor field on \(\scri\). We can extend \(\mu^A\) arbitrarily into the unphysical spacetime to obtain a spinor field \(\tilde \mu^A\), then
\be
    D_b \mu^A \defn \pb{\nabla_b \tilde\mu^A}
\ee
Note that the action of the derivative \(D_b\) is well-defined since it is independent of the extension of \(\mu^A\) into the unphysical spacetime.

 Consider the ``Infeld-van der Waerden symbols'' \(\sigma^a_{AA'}\) in \(M\) which are implicitly used to convert between a tensor index and a pair of spinor indices \cite{PR1}. At \(\scri\), we can express them in our tetrad and spinor basis as
\be
    \sigma^a_{AA'} \hateq n^a \o_A \o_{A'} - m^a \i_A \o_{A'} - \bar m^a \o_A \i_{A'} + l^a \i_A \i_{A'}
\ee
Clearly, \(\sigma^a_{AA'}\) is not intrinsic to \(\scri\). However, lets define
\be\label{eq:Gamma-defn}
    \sigma^a_A &\defn \sigma^a_{AA'} \i^{A'} \eqsp \sigma^a_{A'} &\defn \sigma^a_{AA'} \i^A \, .
\ee
Since these quantities are tangent to \(\scri\), we can consider them as spinor-valued vector fields intrinsically on \(\scri\). By direct computation, they satisfy the identities
\be
    \bar{\sigma^a_A} \hateq \sigma^a_{A'} \eqsp \sigma^a_A \i^A \hateq n^a \eqsp q_{ab} \sigma^a_A \sigma^b_B \hateq 0 \eqsp q_{ab} \sigma^a_A \sigma^b_{B'} \hateq \i_A \i_{B'}
\ee
and their conjugates. Now, we use \(\sigma^a_A\) and \(\sigma^a_{A'}\) to define the spinor-valued derivatives
\be
    D_A \defn \sigma^a_A D_a \hateq \i^{A'} \nabla_{AA'} \eqsp D_{A'} \defn \sigma^a_{A'} D_a \hateq \i^A \nabla_{AA'}
\ee
with \(D_{A'} \hateq \bar{(D_A)}\) and \(\i^A D_A \hateq \i^{A'} D_{A'} \).

If \(\hat D_a\) and \(D_a\) are equivalent derivative operators on \(\scri\) then for any spinor \(\mu_A\) we have
\be\label{eq:D-equiv-spin}
    (\hat D_A - D_A) \mu_B \hateq 0 \eqsp (\hat D_{A'} - D_{A'}) \mu_B \hateq (\hat\rho - \rho) \i_{A'} \i_B \i^C \mu_C
\ee
while, the difference of equivalence classes of derivative operators is given by
\be\label{eq:D-class-spin}
    (\{ \hat D \}_A - \{ D \}_A) \mu_B \hateq (\hat\sigma - \sigma) \i_A \i_B \i^C \mu_C  \eqsp ( \{ \hat D \}_{A'} - \{ D \}_{A'}) \mu_B \hateq 0
\ee
The corresponding action on primed spinors are obtained by taking the complex conjugate of the above equations.

\subsection{BMS vector fields}
\label{sec:bms-vectors}

In this section we recall the definition and properties of BMS vector fields. In general, we will work with a complex BMS vector field \(\xi^a\); the usual real BMS algebra can be obtained using the reality condition \(\bar{\xi^a} = \xi^a\).

A BMS vector field \(\xi^a\) at null infinity can be characterized in the following different ways (see \cite{GPS}), each of which will be useful. Intrinsically on \(\scri\), BMS vector fields preserve the universal structure on \(\scri\) so that
\be\label{eq:bms-int}
    \Lie_\xi q_{ab} \hateq 2 \alpha_{(\xi)} q_{ab} \eqsp \Lie_\xi (A n^a) \hateq - \alpha_{(\xi)} (A n^a)
\ee
for some smooth (complex) function \(\alpha_{(\xi)}\) which depends on the chosen vector field \(\xi^a\). On the other hand, if we view \(\scri\) as the boundary of an unphysical spacetime \((M,g_{ab})\) then\footnote{Note the sign in the second of \cref{eq:bms-ext} is opposite to that of the one used in \cite{GPS} due to our convention \cref{eq:normal-defn}.}
\be\label{eq:bms-ext}
    \Lie_\xi g_{ab} \hateq \alpha_{(\xi)} g_{ab} \eqsp \alpha_{(\xi)} \hateq - \Omega^{-1} (An_a) \xi^a \hateq \tfrac{1}{4} \nabla_a \xi^a
\ee
Note that since \(\xi^a\) is tangent to \(\scri\), \(An_a \xi^a\) vanishes at \(\scri\) and so \(\Omega^{-1}(An_a) \xi^a\) is well-defined there. Finally, in the GHP notation using our choice of basis, a BMS vector field is
\be\label{eq:bms-GHP}
    \xi^a &\hateq (A\beta)n^a + X m^a + \tilde X \bar m^a \\
    \text{with } \thorn'(A\beta) &\hateq \half (\eth X + \eth' \tilde X) \eqsp \thorn' X \hateq \eth'X \hateq 0 \eqsp \thorn' \tilde X \hateq \eth\tilde X \hateq 0
\ee
and the function \(\alpha_{(\xi)}\) is given by
\be\label{eq:alpha-GHP}
    \alpha_{(\xi)} & \hateq \half (\eth X + \eth' \tilde X)
\ee
In \cref{eq:bms-GHP} the GHP weights are
\be
    \beta \wt (0,0) \eqsp X \wt (-1,1) \eqsp \tilde X \wt(1,-1)
\ee
It is easily verified that the BMS vector fields form a Lie algebra \(\bms_{\bb C}\), under the bracket \(\lb[ \xi_1, \xi_2 \rb] \equiv \Lie_{\xi_1} \xi_2^a = \xi^a \in \bms_{\bb C}\). In the GHP notation the Lie brackets can be explicitly written as (\((1 \lra 2)\) indicates the previous terms with the labels \(1\) and \(2\) interchanged)
\be\label{eq:bms-comm}
    \beta & \hateq \beta_1 \eth X_2 - X_2 \eth \beta_1 + \beta_1 \eth' \tilde X_2 - \tilde X_2 \eth' \beta_1 - (1 \lra 2) \\
    X & \hateq X_1 \eth X_2 - X_2 \eth X_1 \\
    \tilde X & \hateq \tilde X_1 \eth' \tilde X_2 - \tilde X_2 \eth' \tilde X_1
\ee
The BMS algebra contains an infinite-dimensional abelian invariant subalgebra (i.e., a Lie ideal) of supertranslations \(\mf s_{\bb C}\) formed by vector fields of the form \((A\beta)n^a\) which are tangent to the null generators of \(\scri\), i.e \(X \hateq \tilde X \hateq 0\). The quotient \(\bms_{\bb C}/\mf s_{\bb C}\) is isomorphic to \(\mf{sl}(2,\bb C) \times \mf{sl}(2,\bb C) \cong \mf{so}_{\bb C}(1,3)\), the complexified Lorentz algebra --- the functions \(X\) and \(\tilde X\) generate each of the \(\mf{sl}(2,\bb C)\) factors. There is also a \(4\)-dimensional Lie ideal of translations \(\mf t_{\bb C} \subset \mf s_{\bb C}\) where the additional condition \(\eth^2(A\beta) \hateq 0\) is satisfied.

The \emph{real} BMS algebra is \(\bms \subset \bms_{\bb C}\) where \(\bar{\xi^a} = \xi^a\) i.e. \(\beta \in \bb R\) and \(\tilde X = \bar X\). There are also two complex subalgebras of \(\bms_{\bb C}\) which we call the \emph{self-dual} and \emph{anti-self-dual} BMS subalgebras, denoted by \(\bms_+\) and \(\bms_-\) respectively. These can be defined in various equivalent ways as follows.
\begin{subequations}\label{eq:bms-sd-defn}\begin{align}
    \xi^a \in \bms_+ \iff
    & ( q_{ab} + i \varepsilon_{ab}) \xi^b \hateq 0 \\
    & \xi^a \hateq \xi^A \i^{A'} \\
    & \xi^a \hateq (A\beta)n^a + X m^a \eqsp \text{ i.e. } \tilde X \hateq 0 
\end{align}\end{subequations}
and similarly,
\begin{subequations}\label{eq:bms-asd-defn}\begin{align}
    \xi^a \in \bms_- \iff 
    & ( q_{ab} - i \varepsilon_{ab}) \xi^b \hateq 0 \\
    & \xi^a \hateq \xi^{A'} \i^A  \\
    & \xi^a \hateq (A\beta)n^a + \tilde X m^a \eqsp  \text{ i.e. } X \hateq 0 
\end{align}\end{subequations}
Note that, in the GHP notation, if \(\xi^a \in \bms_+\) (or \(\xi^a \in \bms_-\)) then we have for the spinors \(\xi^A\) (\(\xi^{A'}\) respectively)
\be\label{eq:bms-spinor}
    \xi^A \hateq (A\beta) \i^A + X \o^A \eqsp \xi^{A'} \hateq (A\beta) \i^{A'} + \tilde X \o^{A'}
\ee
Further, if \(\xi^a \in \bms_+\), then \(\bar{\xi^a} \in \bms_-\), and \(\bms_+ \inter \bms_- = \mf s_{\bb C}\) is the space of supertranslations. It is also straightforward to check that any \(\xi^a \in \bms_{\bb C}\) can be written as a sum \(\xi^a = \xi^a_+ + \xi^a_-\) where \(\xi^a_+ \in \bms_+\) is self-dual and \(\xi^a_- \in \bms_-\) is anti-self-dual. Note this splitting is not unique --- one can add any supertranslation to \(\xi^a_+\) and subtract it from \(\xi^a_-\) without affecting the original vector field. Any real BMS vector field \(\xi^a \in \bms\) can also be split in this (non-unique) way with \(\xi^a_- = \bar{\xi^a_+}\).

\hr

For some computations it will be useful to extend the BMS vector fields away from \(\scri\) into the unphysical spacetime. In general, there is no unique way to do this --- different choices of extensions of the same BMS symmetry correspond to different gauge choices inside the spacetime. However, these extensions are some what restricted as shown in prop.~4.1 of \cite{GPS} which we quote below without proof.
\begin{prop}[Equivalent representatives of a BMS symmetry]
\label{prop:bms-reps}
    If \(\xi^a\) and \({\xi'}^a\) are vector fields in the unphysical spacetime \(M\) which represent the same BMS symmetry at null infinity, i.e. \(\xi^a \hateq {\xi'}^a \in \bms_{\bb C}\), then \({\xi'}^a = \xi^a + O(\Omega^2)\). \\\qed
\end{prop}

Now, let \(\xi^a\) be any smooth vector field in the unphysical spacetime so that \(\xi^a\vert_\scri \in \bms_{\bb C}\). Using \cref{eq:bms-ext} we have that there exists a smooth symmetric tensor \(\gamma_{(\xi)}{}_{ab}\) such that
\be\label{eq:Lie-xi-g}
    \Lie_\xi g_{ab} = 2\alpha_{(\xi)} g_{ab} + \Omega \gamma_{(\xi)}{}_{ab}
\ee
where \(\alpha_{(\xi)}\) is any smooth function away from \(\scri\) subject to the conditions in \cref{eq:bms-ext} at \(\scri\). Further, it can be shown that for the Bondi condition \(\nabla_a \nabla_b \Omega \hateq 0\) to be preserved under the diffeomorphism generated by \(\xi^a\) we need (see \cite{GPS,WZ})
\be\label{eq:gamma-n}
    \gamma_{(\xi)}{}_{ab}(An^b) \hateq 0 \eqsp \gamma_{(\xi)}{}_{AB A'B'}(A \i^B \i^{B'}) \hateq 0
\ee
where the second equation is just the spinor form of the first one. Note that \(\gamma_{(\xi)}{}_{ab} = 0\) if \(\xi^a\) is an exact conformal Killing field in the unphysical spacetime, which will be the case if it is an exact Killing field of the physical spacetime. Thus, BMS vector fields can be viewed as approximate conformal Killing fields of the unphysical spacetime near \(\scri\).

Since \(\Lie_\xi g_{ab} = 2 \nabla_{(a} \xi_{b)}\) we have
\be\label{eq:D-xi}
    \nabla_a \xi_b = \alpha_{(\xi)} g_{ab} + \chi_{(\xi)}{}_{ab} + \half \Omega \gamma_{(\xi)}{}_{ab} \eqsp \chi_{(\xi)}{}_{ab} \defn \nabla_{[a} \xi_{b]}
\ee
Note that it follows from \cref{prop:bms-reps} that the values of \(\alpha_{(\xi)}\) and \(\chi_{(\xi)}{}_{ab}\) at \(\scri\) do not depend on how the BMS vector field was extended into the unphysical spacetime. The values of the tensor \(\gamma_{(\xi)}{}_{ab}\) do depend on the choice of extension of the BMS vector field.

In the spinor notation, it follows from the antisymmetry of \(\chi_{(\xi)}{}_{ab}\) that
\be\label{eq:chi-spinor}
    \chi_{(\xi)}{}_{ab} \hateq -\chi_{(\xi)}{}_{AB} \epsilon_{A'B'} - \tilde\chi_{(\xi)}{}_{A'B'} \epsilon_{AB} \eqsp \chi_{(\xi)}{}^{AB} \hateq - \half \nabla^{(A}_{B'} \xi^{B) B'} \eqsp \tilde\chi_{(\xi)}{}^{A'B'} \hateq - \half \nabla^{(A'}_{B} \xi^{B') B}
\ee
These spinor forms will be useful later (\cref{sec:PRLie}) to define the Lie derivatives of spinor fields. In the GHP notation we have
\be\label{eq:chi-GHP}
    \chi_{(\xi)}{}^{AB}  & \hateq \i^{A} \i^{B} ( - X \sigma - \bar \rho \tilde X  + \eth (A\beta)) + \tfrac{1}{2} \o^{(A} \i^{B)} ( \eth X - \eth' \tilde X ) \\
     \tilde\chi_{(\xi)}{}^{A'B'} & \hateq \i^{A'} \i^{B'} ( - \tilde X \bar \sigma - \rho X  + \eth' (A \beta)) + \tfrac{1}{2} \o^{(A'} \i^{B')} ( \eth' \tilde X - \eth X ) 
\ee
Some other useful identities for the \(\chi_{(\xi)}{}_{ab}\) are proven in \cref{sec:chi-comps}.

\subsection{BMS twistors}
\label{sec:bms-twistor}

In the section we recall the construction of BMS twistors at null infinity from \cite{bms-twistors}. The \emph{BMS twistors} are spinor field solutions \(\omega^A\) on \(\scri\) of the BMS twistor equations which can be expressed in any of the following forms\footnote{Similar equations have been used before in \cite{Helfer:2007ne} but were only imposed on a single null generator of \(\scri\).}
\begin{subequations}\label{eq:bms-twistor}\begin{align}
    \i_B D^{(A} \omega^{B)} \hateq 0 &\eqsp \i_B D^{A'} \omega^B \hateq 0 \label{eq:bms-twistor-int} \\
    \i_{A'} \i_B \nabla^{A'(A} \omega^{B)} \hateq 0 &\eqsp \i_A \i_B \nabla^{A'(A} \omega^{B)} \hateq 0 \label{eq:bms-twistor-ext} \\ 
    \eth' \omega^0 \hateq 0 \eqsp \thorn' \omega^0 \hateq 0 &\eqsp \thorn' \omega^1 \hateq \eth \omega^0 \label{eq:bms-twistor-GHP}
\end{align}\end{subequations}
where in the last expression we have written \(\omega^A = \omega^0 \o^A + \omega^1 \i^A\) in our spinor basis. As explained in \cite{bms-twistors} these equations have an infinite-dimensional space of solutions which we denote by \(\mf T\). We will show in \cref{thm:bms-twistor-inv} that the space of BMS twistors \(\mf T\) is invariant under BMS transformations.

Now let \(\omega^A \equiv (\omega^0 ,\omega^1)\) and \(\tilde\omega^A \equiv (\tilde\omega^0, \tilde\omega^1)\) be any two BMS twistors. Then, we define a self-dual vector field \(\xi^a \in \bms_+ \) on \(\scri\) by
\begin{subequations}\label{eq:twistor-to-vector}\begin{align}
    \xi^a & \hateq 2 i A \sigma^a_A \i_B \omega^{(A} \tilde \omega^{B)} \label{eq:twistor-to-vector-int} \\
    & \hateq \lb( 2 i A \i_B \omega^{(A} \tilde \omega^{B)} \rb)\i^{A'}  \label{eq:twistor-to-vector-ext} \\
    & \hateq (A \beta)n^a + X m^a  \eqsp \beta \hateq - i (\omega^0 \tilde\omega^1 + \omega^1 \tilde\omega^0) \eqsp X \hateq -2i A \omega^0 \tilde\omega^0 \label{eq:twistor-to-vector-GHP} \,.
\end{align}\end{subequations}
Using \cref{eq:bms-twistor} it follows by a direct computation that \(\xi^a\) as defined in \cref{eq:twistor-to-vector} is indeed a self-dual BMS vector field in \(\bms_+\) (see \cite{bms-twistors}).

We point out here that \cref{eq:twistor-to-vector} defines a symmetric map from BMS twistors \(\mf T\) to the self-dual BMS vector fields \(\bms_+\). This structure is very reminiscent of an anticommutator in a superalgebra; we will shown in \cref{sec:superalgebra}. that this is indeed the case.

\section{Lie derivative of spinors along BMS vector fields}
\label{sec:PRLie}

Since our ultimate goal is to obtain a superalgebra combining the BMS vector fields and the BMS twistors, we need to define an action of the BMS symmetries on spinor fields. In general, the Lie derivative of a spinor field along a given vector field is not uniquely defined; the reason being that spinor fields depend not only on the spacetime manifold but also on a choice of local frames, that is they sections of a vector bundle associated to the frame bundle \cite{Lich,Kos,FF,GM,LRW,Helfer,Prabhu:2015vua}. It was shown by Habermann \cite{Habermann} that there is a good notion of Lie derivative of a spinor field along conformal Killing fields which interacts nicely with the (full) twistor equation --- this Lie derivative is the same as the one defined by Penrose and Rindler in \S~6.6 of \cite{PR2}. Since at \(\scri\), the BMS vector fields are indeed conformal Killing fields (see \cref{eq:bms-ext}), we will use the Penrose-Rindler definition of the Lie derivative along BMS vector fields.

For any \(\xi^a \in \bms_{\bb C}\), define the Lie derivative of spinor fields \(\mu^A\) and \(\nu^{A'}\) on \(\scri\) by
\be\label{eq:Lie-spinor}
    \Lie_\xi \mu^A &\defn \xi^b \nabla_b \mu^A + \Xi_{(\xi)}{}^A{}_B \mu^B \eqsp && \Xi_{(\xi)}{}^A{}_B ~\hateq \chi_{(\xi)}{}^A{}_B - \tfrac{1}{2} \alpha_{(\xi)} \epsilon_B{}^A \\
    \Lie_\xi \nu^{A'} &\defn \xi^b \nabla_b \nu^{A'} + \tilde\Xi_{(\xi)}{}^{A'}{}_{B'} \nu^{B'} \eqsp && \tilde\Xi_{(\xi)}{}^{A'}{}_{B'} \hateq \tilde\chi_{(\xi)}{}^{A'}{}_{B'} - \tfrac{1}{2} \alpha_{(\xi)} \epsilon_{B'}{}^{A'} 
\ee
This Lie derivative can be extended to spinor fields with many indices, primed or unprimed, in the usual way. By direct computation we have (see also \cite{Helfer}, with \(w  = -1\))
\be\label{eq:Lie-epsilon-sigma}
    \Lie_\xi \epsilon_{AB} \hateq \alpha_{(\xi)} \epsilon_{AB} \eqsp \Lie_\xi \epsilon_{A'B'} \hateq \alpha_{(\xi)} \epsilon_{A'B'} \eqsp \Lie_\xi \sigma^a{}_{AA'} \hateq 0
\ee
The first two equations imply that we can raise and lower spinor indices inside a \(\Lie_\xi\) by compensating with appropriate terms involving \(\alpha_{(\xi)}\), while the last equation implies that we are allowed to convert from spinor indices \(AA'\) to the tensor index \(a\) inside a \(\Lie_\xi\).

It will be useful to have the following identities at hand. Writing the second condition in \cref{eq:bms-int} in terms of \(n^a \hateq \i^A \i^{A'}\) we get
\be
    (\Lie_\xi A) \i^A \i^{A'} + A (\Lie_\xi \i^A) \i^{A'} + A \i^A \Lie_\xi \i^{A'} \hateq - \alpha_{(\xi)} (A \i^A \i^{A'})
\ee
which implies
\be\label{eq:Lie-i}
    \Lie_\xi \i^A \propto \i^A \eqsp \Lie_\xi \i^{A'} \propto \i^{A'}
\ee
Similarly,
\be\label{eq:Lie-sigma-i}
    \Lie_\xi (A \sigma^a_A \i_B) \hateq \Lie_\xi (A \sigma^a_{AA'} \i^{A'} \i_B) \hateq \Lie_\xi (A \sigma^a_{AA'}   \sigma_b{}^{A'}{}_B n^b) \hateq 0
\ee
Finally, for \(\xi^a \in \bms_{\bb C}\) and any spinor field \(\mu^A\), we have (see \cite{Helfer}, with \(w = -1\))
\be\label{eq:Lie-D-comm}
    \lb[ \Lie_\xi, \nabla_b \rb] \mu^A &\hateq \xi^d R_{db C}{}^A \mu^C - ( \nabla_b \chi_{(\xi)}{}_C{}^A ) \mu^C + \tfrac{1}{2} (\nabla_b \alpha_{(\xi)}) \mu^A \\
    & \hateq \half \lb(\epsilon_{BC} \nabla_{B'}^A \alpha_{(\xi)} + \epsilon_B{}^A \nabla_{CB'}\alpha_{(\xi)} \rb) \mu^C + \half (\nabla_{BB'} \alpha_{(\xi)}) \mu^A \\
    &\quad + \nfrac{1}{4} A \lb( \i_{C} \gamma_{(\xi)}{}^A{}_{BC'B'} + \i^A \gamma_{(\xi)}{}_{CBC'B'} \rb) \i^{C'} \mu^C
\ee
where the second line uses \cref{eq:D-chi}.

In the GHP notation the Lie derivative of a spinor field \(\mu^A = \mu^0 \o^A + \mu^1 \i^A\) takes the following form (using \cref{eq:bms-GHP})
\be\label{eq:Lie-spinor-GHP}
    \Lie_\xi \mu^A &\hateq \o^A \lb[ (A\beta) \thorn' \mu^0 + X \eth \mu^0 - \tfrac{1}{2} \mu^0 \eth X + \tilde X \eth'\mu^0 \rb] \\
    &\quad~ + \i^A \lb[ (A\beta) \thorn' \mu^1 - \mu^0 \eth(A\beta) + X \eth\mu^1 + \tilde X \eth' \mu^1 - \tfrac{1}{2} \mu^1 \eth' \tilde X \rb]
\ee
One can also compute a similar expression for primed spinor fields. When \(\mu^A = \omega^A \in \mf T\) is a BMS twistor \cref{eq:Lie-spinor-GHP} simplifies to (using \cref{eq:bms-twistor-GHP})
\be\label{eq:Lie-twistor-GHP}
    \Lie_\xi \omega^A &\hateq \o^A \lb[ X \eth \omega^0 - \tfrac{1}{2} \omega^0 \eth X \rb] \\
    &\quad~ + \i^A \lb[ (A\beta) \eth\omega^0 - \omega^0 \eth(A\beta) + X \eth\omega^1 + \tilde X \eth' \omega^1 - \tfrac{1}{2} \omega^1 \eth' \tilde X \rb]
\ee

Now, note that since \(\xi^a \in \bms_{\bb C}\) is tangent to \(\scri\), in \cref{eq:Lie-spinor} we can replace \(\nabla_a\) by its pullback, the intrinsic derivative operator \(D_a\) on \(\scri\). Next we notice that the non-universal spin coefficients \(\Re\rho\) and \(\sigma\) do not appear in \cref{eq:Lie-spinor-GHP} --- the terms containing these spin coefficients exactly cancel between \(\xi^b \nabla_b \mu^A\) and the \(\chi_{(\xi)}{}_{ab}\) in \cref{eq:Lie-spinor}. Thus, we have the following result
\begin{lemma}\label{lem:Lie-univ}
For any \(\xi^a \in \bms_{\bb C}\) the Lie derivative of spinor fields on \(\scri\) defined in \cref{eq:Lie-spinor} is intrinsic and universal, i.e., independent of the choice of derivative operator \(D_a\) on \(\scri\). \\\qed
\end{lemma}

Next, we show that the Lie derivative \cref{eq:Lie-spinor} gives a well-defined action of the BMS Lie algebra on spinor fields.

\begin{lemma}\label{lem:Lie-comm}
For \(\xi_1^a, \xi_2^a \in \bms_{\bb C}\), and any spinor field \(\mu^A\) on \(\scri\) we have
\be\label{eq:Lie-comm}
    [ \Lie_{\xi_1},  \Lie_{\xi_2}] \mu^A \hateq \Lie_{[\xi_1 , \xi_2]} \mu^A
\ee
\begin{proof}
    Let \(\xi^a \equiv [\xi_1, \xi_2]\). Then we have from \cref{eq:Lie-spinor}
    \be
    \lb[ \Lie_{\xi_1}, \Lie_{\xi_2} \rb] \mu^A & \hateq \xi^b \nabla_b \mu^A + \Lie_{\xi_1} \Xi_{(\xi_2)}{}^A{}_B \mu^B + \xi_2^b \lb[ \Lie_{\xi_1}, \nabla_b \rb] \mu^A
    \ee
    Then using \cref{eq:Lie-spinor,eq:Lie-D-comm} we get
    \be
    \lb( \lb[ \Lie_{\xi_1}, \Lie_{\xi_2} \rb] - \Lie_\xi \rb) \mu^A & \hateq - \Xi_{(\xi)}{}^A{}_B \mu^B + \Lie_{\xi_1} \Xi_{(\xi_2)}{}{}^A{}_B \mu^B + \xi_2^b \lb( \xi_1^c R_{cb D}{}^A - \nabla_b \Xi_{(\xi_1)}{}{}^A{}_D \rb) \mu^D \\
    & \hateq \big[ - \Xi_{(\xi)}{}^A{}_B + \xi_1^c \nabla_c \Xi_{(\xi_2)}{}{}^A{}_B - \xi_2^c \nabla_c \Xi_{(\xi_1)}{}{}^A{}_B \\
    &\qquad + \Xi_{(\xi_1)}{}{}^A{}_C \Xi_{(\xi_2)}{}{}^C{}_B - \Xi_{(\xi_2)}{}{}^A{}_C \Xi_{(\xi_1)}{}{}^C{}_B + \xi_1^c \xi_2^d R_{cd B}{}^A \big] \mu^B
\ee
From \cref{prop:alpha-chi-Lie} the right-hand-side vanishes, and we have the desired result.
\end{proof}
\end{lemma}

Next we prove one of our main results --- the space \(\mf T\) of BMS twistors is invariant under the action of the complex BMS algebra \(\bms_{\bb C}\). One can prove this directly using the GHP expressions \cref{eq:bms-twistor-GHP,eq:Lie-twistor-GHP}, but this obscures the crucial role played by the properties of BMS vector fields and the BMS twistor equations. So in the following we provide a covariant proof and the GHP computation is collected in \cref{rem:bms-twistor-inv-GHP}.

\begin{thm}
\label{thm:bms-twistor-inv}
Let \(\xi^a \in \bms_{\bb C}\) be a complex BMS vector field and \(\omega^A \in \mf T\) be a BMS twistor then \(\Lie_\xi \omega^A \in \mf T\) is also a BMS twistor.
\begin{proof}
For a BMS vector field \(\xi^a \in \bms_{\bb C}\), using \cref{eq:Lie-D-comm}, we have
\be
    \lb[ \Lie_\xi, \nabla_b \rb] \omega^A 
    & \hateq \half \lb(\epsilon_{BC} \nabla_{B'}^A \alpha_{(\xi)} + \epsilon_B{}^A \nabla_{CB'}\alpha_{(\xi)} \rb) \omega^C + \half (\nabla_{BB'} \alpha_{(\xi)}) \omega^A \\
    &\quad + \nfrac{1}{4} A \lb( \i_{C} \gamma_{(\xi)}{}^A{}_{BC'B'} + \i^A \gamma_{(\xi)}{}_{CBC'B'} \rb) \i^{C'} \omega^C
\ee
Next, raise \(b\) index, convert it to \(BB'\) and symmetrize in the indices \(A\) and \(B\) to get
\be\label{eq:Lie-twistor-comm}
    \lb[ \Lie_\xi, \nabla^{B'(B} \rb] \omega^{A)} \hateq - 2 \alpha_{(\xi)} \nabla^{B'(B} \omega^{A)} + \tfrac{1}{4} A  \lb( \gamma_{(\xi)}{}^{ABB'}{}_{C'} \i_C + \i^{(A} \gamma_{(\xi)}{}^{B)B'}{}_{CC'}\rb)\i^{C'} \omega^C
\ee
where the first term on the right-hand-side comes from raising the \(b\) index using the metric. Next, we contract the above equation with \(\i_A\) and \(\i_{A'}\), and use \cref{eq:gamma-n} to get
\be
    \i_A \i_{B'} \lb[ \Lie_\xi, \nabla^{B'(B} \rb] \omega^{A)}  & \hateq - 2 \alpha_{(\xi)} \i_A \i_{B'} \nabla^{B'(B} \omega^{A)} \\
    \i_A \i_B \lb[ \Lie_\xi, \nabla^{B'(B} \rb] \omega^{A)} & \hateq  - 2 \alpha_{(\xi)} \i_A \i_B \nabla^{B'(B} \omega^{A)}
\ee
Then, using the fact that \(\Lie_\xi \i^A \propto \i^A\) (\cref{eq:Lie-i}) we have that \(\Lie_\xi \omega^A\) satisfies the BMS twistor equations (in the form \cref{eq:bms-twistor-ext}) whenever \(\omega^A\) does, that is, the BMS twistor equations are preserved under the action of \(\bms_{\bb C}\).
\end{proof}
\end{thm}

Note that if \(\xi^a\) is an exact conformal Killing vector then \(\gamma_{(\xi)}{}_{ab} = 0\), and so \cref{eq:Lie-twistor-comm} implies that the solution space of the full twistor equation is preserved, reproducing the result of Habermann \cite{Habermann}. But, for a general BMS vector field \(\gamma_{(\xi)}{}_{ab} \neq 0\) and so the full twistor equation is not preserved under BMS symmetries; it is only the BMS twistor equations which are preserved.

\begin{remark}[GHP proof of \cref{thm:bms-twistor-inv}]
\label{rem:bms-twistor-inv-GHP}
Let \(\tilde \omega^A = \Lie_\xi \omega^A\), from \cref{eq:Lie-twistor-GHP}, we have
\be
    \tilde \omega^0 = X \eth \omega^0 - \tfrac{1}{2} \omega^0 \eth X  \eqsp \tilde \omega^1 = (A\beta)\eth\omega^0 - \omega^0 \eth(A\beta) + X \eth \omega^1  + \tilde X \eth' \omega^1 - \tfrac{1}{2} \omega^1 \eth' \tilde X 
\ee
Now we would like to verify that \(\tilde \omega^A\) satisfies the BMS twistor equations (in the GHP form \cref{eq:bms-twistor-GHP}) whenever \(\omega^A\) does. For this we use the following commutators of the GHP derivatives at \(\scri\)
\be\label{eq:GHP-comm}
    \lb[\thorn', \eth'\rb] \eta = \lb[\thorn', \eth\rb] \eta = 0  \eqsp \lb[ \eth, \eth' \rb]\eta = - s \ms R \eta
\ee
where \(s\) is the spin of the field \(\eta\) and \(\ms R\) is the Ricci scalar of the induced metric \(q_{ab}\).
Then, using \cref{eq:bms-GHP,eq:bms-twistor-GHP}, one can easily check \(\thorn' \tilde\omega^0 = 0\). Next, we have
\be
    \eth' \tilde\omega^0 = X \eth' \eth \omega^0 - \tfrac{1}{2} \omega^0 \eth' \eth X = X(-\half \ms R \omega^0) - \tfrac{1}{2}\omega^0 (- \ms R X) = 0 
\ee
where we have used \(\omega^0\) is spin \(s = -\half\) and \(\eth'\omega^0 = 0\), and \(X\) is spin \(s = -1\) and \(\eth' X= 0\). Finally, we compute
\be
    \thorn' \tilde\omega^1 & = \thorn'(A\beta) \eth\omega^0 - \omega^0 \eth \thorn' (A\beta) +  X \eth \thorn' \omega^1 + \tilde X \eth' \thorn' \omega^1 - \tfrac{1}{2} \thorn'\omega^1 \eth' \tilde X \\
    & = X \eth^2 \omega^0 + \half \eth X \eth\omega^0 - \half \omega^0 \eth^2 X - \half \omega^0 \eth \eth'\tilde X + \tilde X \eth' \eth \omega^0 \\
    & = \eth \tilde \omega^0  - \half \omega^0 (- \ms R \tilde X) + \tilde X (-\half \ms R \omega^0) = \eth \tilde \omega^0
\ee
where in the second line we have used \cref{eq:bms-GHP}. The third line uses \cref{eq:GHP-comm} and that \(\omega^0\) is spin \(s = \half\) and \(\eth'\omega^0 = 0\), and \(\tilde X\) is spin \(s = 1\) and \(\eth \tilde X = 0\).
\end{remark}

\section{The BMS Lie superalgebra \(\mf K\)}
\label{sec:superalgebra}

In this section we combine the self-dual BMS vector fields and the BMS twistors into a Lie superalgebra which we denote by \(\mf K\).\footnote{A similar construction can be carried out with the anti-self-dual BMS vector fields \(\bms_-\) and the complex conjugate BMS twistors \(\bar{\mf T}\) to get the Lie superalgebra \(\bar{\mf K} = \bms_- \oplus \bar{\mf T}\).} We note here that our construction differs from other supersymmetric extensions of the BMS algebra considered previously in the context of supergravity --- we explain these differences in \cref{rem:compare}.

We consider the supervector space, i.e., a \(\bb Z_2\)-graded vector space
\be
    \mf K \defn \bms_+ \oplus \mf T
\ee
where \(\bms_+\) is assigned an even grading and \(\mf T\) is assigned an odd grading. Next we define a graded bracket on \(\mf K\) by
\begin{subequations}\label{eq:brackets}\begin{align}
    \lb[\xi_1, \xi_2\rb] = - \lb[\xi_2, \xi_1\rb] & \equiv \Lie_{\xi_1} \xi_2^a \in \bms_+ \label{eq:bracket00} \\
    \lb[ \omega_1, \omega_2 \rb] = \lb[ \omega_2, \omega_1 \rb] &\equiv 2iA \sigma^a_A \i_B \omega_1^{(A} \omega_2^{B)} \in \bms_+ \label{eq:bracket11} \\
    \lb[\xi, \omega\rb] = - \lb[ \omega, \xi \rb] &\equiv \Lie_\xi \omega^A \in \mf T \label{eq:bracket01}
\end{align}\end{subequations}
\Cref{eq:bracket00} is just the Lie bracket of the \(\bms_+\) algebra, \cref{eq:bracket11} defines a symmetric bracket on \(\mf T\) which returns the self-dual BMS vector field formed from the two BMS twistors according to \cref{eq:twistor-to-vector-int}, and \cref{eq:bracket01} is the Lie derivative along a BMS vector field of the twistor \(\omega^A\) which returns another BMS twistor as shown in \cref{thm:bms-twistor-inv}.

Taking the brackets defined in \cref{eq:brackets} as a graded product gives the supervector space \(\mf K\) the structure of a superalgebra. However, for \(\mf K\) to be a \emph{Lie} superalgebra the graded Jacobi identities must be satisfied which we prove in the following theorem. We emphasize that the satisfaction of the Jacobi identities is quite non-trivial. For instance, if we replace \(\bms_+\) by the algebra of exact conformal Killing fields and \(\mf T\) by solutions of the full twistor equation, then the Jacobi identities are not satisfied in general as shown by Habermann \cite{Habermann}.

\begin{thm}
    The supervector space \(\mf K = \bms_+ \oplus \mf T\), with \(\bms_+\) the even subspace and \(\mf T\) the odd subspace, equipped with graded brackets defined by \cref{eq:brackets}, is a Lie superalgebra.
\begin{proof}
For our case, the four linearly-independent Jacobi identities, which we need to verify, are
\be\label{eq:jacobi}
    0 & = + \lb[ \xi_1 , \lb[\xi_2, \xi_3\rb] \rb] + \lb[ \xi_2 , \lb[\xi_3, \xi_1\rb] \rb] + \lb[ \xi_3 , \lb[\xi_1, \xi_2\rb] \rb] \\
    0 & = + \lb[ \xi_1 , \lb[\xi_2, \omega\rb] \rb] + \lb[ \xi_2 , \lb[\omega, \xi_1\rb] \rb] + \lb[ \omega , \lb[\xi_1, \xi_2\rb] \rb] \\
    0 & = + \lb[ \omega_1 , \lb[\omega_2, \xi \rb] \rb] - \lb[ \omega_2 , \lb[\xi, \omega_1\rb] \rb] + \lb[ \xi , \lb[\omega_1, \omega_2\rb] \rb] \\
    0 & = - \lb[ \omega_1 , \lb[\omega_2, \omega_3\rb] \rb] - \lb[ \omega_2 , \lb[\omega_3, \omega_1\rb] \rb] - \lb[ \omega_3 , \lb[\omega_1, \omega_2\rb] \rb]
\ee
The first of these is just the Jacobi identity for the Lie algebra \(\bms_+\), while the second follows from \cref{lem:Lie-comm}. Thus, we only need to check the last two Jacobi identities, which we prove by direct computation as follows.

Consider
\be
    \lb[ \omega_1, \lb[ \omega_2 ,\xi \rb] \rb] = - \lb[ \omega_1 , \Lie_\xi \omega_2 \rb] &\equiv - 2  i A \sigma^a_A \i_B \omega_1^{(A} \Lie_\xi \omega_2^{B)} \\
    & = \Lie_\xi \lb(- 2  i A \sigma^a_A \i_B \omega_1^{(A} \omega_2^{B)} \rb) + 2  i A \sigma^a_A \i_B \Lie_\xi \omega_1^{(A} \omega_2^{B)} \\
    & \equiv - \lb[ \xi, \lb[ \omega_1 ,\omega_2\rb] \rb] + \lb[ \lb[\xi, \omega_1\rb] , \omega_2 \rb]
\ee
where the second line uses the Leibniz rule and \cref{eq:Lie-sigma-i}. Rearranging the above equation proves the third Jacobi identity.

Next, let \([\omega_2,\omega_3] \equiv \xi^a = (A\beta) n^a + X m^a \in \bms_+\) and consider
\be
    - [\omega_1, [\omega_2, \omega_3]] & \equiv \Lie_\xi \omega_1^A \\
    & = \o^A \lb[ X \eth \omega_1^0 - \tfrac{1}{2} \omega_1^0 \eth X \rb] + \i^A \lb[ X \eth \omega_1^1 + (A\beta)\eth \omega_1^0 - \omega_1^0 \eth(A\beta) \rb] \\
    & = i A \o^A \lb[ -2 (\eth \omega_1^0) \omega_2^0 \omega_3^0 + \omega_1^0 (\eth \omega_2^0) \omega_3^0 + \omega_1^0 \omega_2^0 (\eth\omega_3^0)  \rb] \\
    &\quad + iA \i^A \lb[ -2 (\eth \omega_1^1) \omega_2^0 \omega_3^0 - (\eth \omega_1^0) (\omega_2^0 \omega_3^1 + \omega_2^1 \omega_3^0)  + \omega_1^0 \eth (\omega_2^0 \omega_3^1 + \omega_2^1 \omega_3^0) \rb]
\ee
where in the second line we have used \cref{eq:Lie-twistor-GHP} (and \(\tilde X = 0\) for a self-dual BMS vector field) and in the third line we use \cref{eq:twistor-to-vector-GHP} to write the vector field \(\xi^a\) in terms of the BMS twistors \(\omega_2^A\) and \(\omega_3^A\). Substituting this, and similar equations obtained by a cyclic permutation of the three BMS twistors, into the last Jacobi identity we see that all the terms cancel, and the last Jacobi identity is satisfied.
\end{proof}
\end{thm}

\begin{remark}[Other supersymmetric extensions of BMS]\label{rem:compare}
In the context of supergravity, supersymmetric extensions of the BMS algebra have been investigated at null infinity by Awada, Gibbons and Shaw \cite{AGS} (see also \cite{Fuentealba:2021xhn} for an analysis at spatial infinity). The construction of \cite{AGS} can be summarized as follows. One attempts to impose all the components of the twistor equation which are tangent to \(\scri\); these are
\be
    \thorn' \omega^0 = 0 \eqsp \eth'\omega^0 = 0 \eqsp \thorn'\omega^1 = \eth \omega^0 \eqsp \eth\omega^1 = \sigma \omega^0
\ee
Note that the first three are the BMS twistor equations (\cref{eq:bms-twistor-GHP}), but the last one depends on the shear \(\sigma\) and is not universal. It is well-known that this set of equations has no non-trivial solution unless \(\bar N = 0\) (i.e. there is no radiation), or \(\omega^0 = 0\) \cite{DS,AGS,bms-twistors}. Since we definitely want to consider spacetimes with radiation at \(\scri\), we choose to impose \(\omega^0 = 0\). Then using \(\omega^1\) and its complex conjugate \(\bar\omega^{1'}\) we can form a BMS vector field
\be\label{eq:AGS-bms}
    \xi^a = - i (\omega^1 \bar\omega^{1'}) n^a
\ee
In general this is a BMS supertranslation, and if we impose \(\eth \omega^1 = 0\) we get a BMS translation. Note one can also generate complex supertranslations by replacing \(\bar\omega^{1'}\) by a second solution \(\bar{\tilde \omega}^{1'}\). Then, one can construct a superalgebra by defining graded brackets similar to \cref{eq:brackets} but the bracket of two twistors is replaced by \(\lb[ \omega ,\bar\omega \rb] \equiv \xi^a\) where \(\xi^a\) is the BMS supertranslation in \cref{eq:AGS-bms}. The differences with our construction of the superalgebra \(\mf K\) are quite apparent. As detailed in \cite{bms-twistors} we only impose the components of the twistor equation which are intrinsic and universal, i.e. we do not impose \(\eth \omega^1 = \sigma \omega^0\), and consequently have a non-zero \(\omega^0\). It is precisely this non-zero \(\omega^0\) which helps us generate the entire (self-dual) BMS algebra including Lorentz vector fields, unlike the construction of \cite{AGS,Fuentealba:2021xhn} which only generates supertranslations. As we will describe in \cref{sec:NS}, this \(\omega^0\) can be viewed as generating Neveu-Schwarz-type supersymmetries on \(\bb S^2\), which do not appear in the construction of \cite{AGS,Fuentealba:2021xhn}.
\end{remark}

\subsection{Projection to the Neveu-Schwarz superalgebra on \(\bb S^2\)}
\label{sec:NS}

In this section we show how the BMS Lie superalgebra \(\mf K\) projects to the global Neveu-Schwarz (NS) superalgebra on a \(2\)-sphere. The NS superalgebra is a supersymmetric extension of the Virasoro algebra of \emph{local} conformal Killing fields of the \(2\)-sphere metric, and it is only the Lorentz subalgebra of the Virasoro algebra which extends to globally smooth conformal Killing fields on a \(2\)-sphere. Since we want the smoothness structure of null infinity to be preserved we restrict to the Lorentz subalgebra of the full Virasoro algebra, and correspondingly restrict to the globally smooth supersymmetric extension.

 To obtain the desired projection map we note that there is a projection \(\scri \to \bb S^2\) which maps every point \(p\) on \(\scri\) to a unique point on \(\bb S^2\) representing the null generator which contains \(p\). The pushforward of this projection maps vector fields on \(\scri\) to vector fields on \(\bb S^2\), in particular, it maps \(\bms_+\) to the quotient algebra \(\bms_+/\mf s_{\bb C} \cong \mf{sl}(2,\bb C)\). Explictly, the projection acts on \(\xi^a \in \bms_+\) as
\be
    \xi^a = (A\beta)n^a + X m^a \mapsto X m^a
\ee
where we recall that \(X\) is a function on \(\bb S^2\) with spin \(s = -1\) and \(\eth'X = 0\). On the space of BMS twistors the projection map acts as
\be
    \omega^A = \omega^0 \o^A + \omega^1 \i^A \mapsto - \i_A \omega^{A} = \omega^0
\ee
where \(\omega^0\) is a function on \(\bb S^2\) with spin \(s = -\half\) and \(\eth' \omega^0 = 0\). Applying the projection to the brackets (\cref{eq:brackets}) we get
\be\label{eq:proj-brackets}
    \lb[ X_1, X_2 \rb] = X_1 \eth X_2 - X_2 \eth X_1 \eqsp \lb[ X, \omega^0 \rb] = X \eth \omega^0 - \tfrac{1}{2} \omega^0 \eth X  \eqsp \lb[ \omega_1^0, \omega_2^0 \rb] = -2i \omega_1^0 \omega_2^0
\ee
We have set the function \(A=1\) since we do not have to consider GHP weights on \(\bb S^2\); only the spin weights will be important which are unaffected by the function \(A\). It can be checked that the Jacobi identities are still satisfied after the projection. Note that the projection maps an infinite-dimensional superalgebra on \(\scri\) into a finite-dimensional one on \(\bb S^2\), just like it maps the BMS algebra into the Lorentz algebra.

We claim that the Lie superalgebra defined by \cref{eq:proj-brackets} is precisely the global subalgebra of the NS supersymmetries.\footnote{The restriction to the global NS subalgebra arises because we want all the fields to be smooth on \(\bb S^2\), which is dictated by the smoothness of \(\scri\). We also note that the central extension plays no role when restricted to the global NS subalgebra.} Since, the NS superalgebra is usually presented in a choice of basis, we need to pick a basis for the functions \(X\) and \(\omega^0\) on \(\bb S^2\) to show this equivalence. In \(2\)-dimensional conformal field theory and string theory it is conventional to work on the complex plane which maps conformally to a \(2\)-sphere using complex stereographic coordinates \((z, \bar z)\). However, the topology of the space of generators of \(\scri\) being a \(2\)-sphere is crucial for asymptotic flatness. One can still work in the stereographic coordinates, if desired, but then one needs to impose suitable regularity conditions on the fields at the ``point at infinity'' \(z = \infty\) to ensure that they define smooth fields on \(\bb S^2\). Instead, it is convenient to choose a conformal factor so that the metric on the space of generators of \(\scri\) is the standard unit metric on \(\bb S^2\), and use spin-weighted spherical harmonics \(Y^s_{\ell,m}\) (which are smooth everywhere on \(\bb S^2\)) to define the required basis. We detail the computation in \cref{sec:NS-comp}, and summarize the main results below.

Note that \(X\) being \(s = -1\) and \(\eth' X = 0\) implies that \(X\) is \(\ell = 1\). Similarly, \(\omega^0\) being \(s = -\half\) and \(\eth' \omega^0 = 0\) implies \(\omega^0\) is \(\ell = \half\). Thus, we define a basis \(L_n\) and \(G_r\), for the space of \(X\) and \(\omega^0\) respectively, by
\be\label{eq:NS-basis}
    L_n \defn - \lb(2^{\abs n} \tfrac{4\pi}{3} \rb)^\half Y^{s=-1}_{\ell=1,m=n} \eqsp G_r \defn \lb( -i 2 \sqrt{2} \pi \rb)^\half Y^{s=-\half}_{\ell = \half, m=r }
\ee
where \(n \in \{-1,0,1\}\) and \(r \in \{-\half,\half\}\). The overall factors have been chosen to bring the final brackets into the same form as the standard NS superalgebra. Next, we substitute the \(X\)s and \(\omega^0\)s in \cref{eq:proj-brackets} by the corresponding basis elements \(L_n\) and \(G_r\). A straightforward computation using the properties of the spin-weighted spherical harmonics (see \cref{sec:NS-comp}) gives the following brackets on the basis elements
\be\label{eq:NS-brackets}
    \lb[ L_n, L_{n'} \rb] = (n - n') L_{n+n'} \eqsp \lb[ L_n, G_r \rb] = (\tfrac{n}{2}-r) G_{n+r} \eqsp \lb[ G_r, G_{r'} \rb] = 2 L_{r+r'} 
\ee
These are precisely the brackets of the global NS superalgebra, see \S~4.2 of \cite{GSW} or \cite{FMS}.

\section{Discussion}
\label{sec:disc}

Lets us point out there there are three distinct kinds of ``superness'' involved in the Lie superalgebra \(\mf K\): the first is the usual extension of the Poincar\'e algebra to the infinite-dimensional BMS algebra, the second is a similar extension of the \(2\)-surface twistors \cite{Penrose-charges, DS, Shaw} to the infinite-dimensional space of BMS twistors defined in \cite{bms-twistors}, and lastly, we have the supersymmetric structure which combines both the BMS algebra and the BMS twistor space described in this paper.

While we have defined the supersymmetric algebra \(\mf K\) on null infinity, we have not specified the action of these supersymmetries on physical fields. It would be of interest to define an action of these supersymmetries on the asymptotic radiative fields, for instance the News tensor which characterizes the asymptotic gravitational radiation. In particular, the relation of the superalgebra \(\mf K\) to bulk supersymmetries in supergravity (if there is any) is unclear at present --- as noted in \cref{rem:compare} any such relation to bulk supersymmetries cannot be the same as the one used in \cite{AGS,Fuentealba:2021xhn}.

It would also be of interest to obtain a superspace formulation of the superalgebra \(\mf K\). In this context we note that the projection of \(\mf K\) to the \(2\)-sphere, that is, the global NS superalgebra (as described in \cref{sec:NS}) does have a superspace formulation \cite{FMS}. A suitable lift of this supermanifold structure from \(\bb S^2\) to \(\scri\) should give a superspace description of the algebra \(\mf K\). We also note that the universal geometric structure of null infinity is a \emph{conformal Carroll structure} on \(\scri\), which is an ``ultrarelativistic'' limit (speed of light tends to zero) of conformal Lorentzian structures \cite{LL,Duval:2014uva,Duval:2014lpa,Hartong:2015xda,Ciambelli:2019lap,Figueroa-OFarrill:2019sex}. In this sense, the appropriate superspace structure on \(\scri\) could be a ``conformal super-Carrollian space'' which, as far as we know, has not been investigated.

\section*{Acknowledgements}
I would like to thank Gary Gibbons and Gautam Satishchandran for suggesting a possible supersymmetric interpretation of BMS twistors which motivated this work. This work is supported by NSF grant PHY-2107939. Some calculations in this paper used the computer algebra system \textsc{Mathematica} \cite{Mathematica}, in combination with the \textsc{xAct/xTensor} suite~\cite{xact,xact-spinors}.

\appendix

\section{Some useful computations with BMS vector fields}
\label{sec:chi-comps}

In the following we collect some side computations with the tensors \(\alpha_{(\xi)}\) and \(\chi_{(\xi)}{}_{ab}\) for BMS vector fields. Since BMS vector fields are approximate conformal Killing fields on the unphysical spacetime (see \cref{eq:Lie-xi-g}) the derivative of \(\chi_{(\xi)}{}_{ab}\), can be written in terms of the Riemann tensor at \(\scri\). The proof is very similar to the case of Killing vector fields given in \S~C.3 of \cite{Wald-book}; and for exact conformal Killing fields one can just set \(\gamma_{(\xi)}{}_{ab} = 0\) in the following results without any need to evaluate at null infinity.
\begin{prop}
\label{prop:D-chi}
\begin{subequations}\begin{align}
    \nabla_a \chi_{(\xi)}{}_{bc} & \hateq - \xi^d R_{dabc} - 2 g_{a[b} \nabla_{c]} \alpha_{(\xi)} + A \gamma_{(\xi)}{}_{a[b} n_{c]}    \label{eq:D-chi-tensor} \\
    -\nabla_a \chi_{(\xi)}{}_{BC} & \hateq - \xi^d R_{daBC} + \epsilon_{A(B} \nabla_{C)A'}\alpha_{(\xi)} + \half A \i^{C'} \gamma_{(\xi)}{}_{A'C'A(B}\i_{C)} \label{eq:D-chi}
\end{align}\end{subequations}
\begin{proof}
We start with the definition of the Riemann tensor (\cref{eq:Riem-defn}):
\be
    \nabla_a \nabla_b \xi_c - \nabla_b \nabla_a \xi_c = - R_{abc}{}^d \xi_d
\ee
In the second term on the left-hand-side we replace \(\nabla_a \xi_c\) by \(-\nabla_c \xi_a + 2 \nabla_{(a} \xi_{c)}\) and then replace \(2 \nabla_{(a} \xi_{c)}\) using \cref{eq:Lie-xi-g}. Evaluating the resulting expression at \(\scri\) we get
\be
    \nabla_a \nabla_b \xi_c + \nabla_b \nabla_c \xi_a  \hateq - R_{abc}{}^d \xi_d  + 2 \nabla_b \alpha_{(\xi)}{} g_{ac} + \nabla_b \Omega \gamma_{(\xi)}{}_{ac}
\ee
Taking cyclic permutations of the indices \(a,b,c\) we get two more equations. Adding two of these equations and subtracting the third we get
\be
    \nabla_a \nabla_b \xi_c \hateq R_{bca}{}^d \xi_d + \lb( \nabla_a\alpha_{(\xi)}{} g_{bc} + \nabla_b \alpha_{(\xi)}{} g_{ca} - \nabla_c \alpha_{(\xi)}{} g_{ba}\rb) - \tfrac{1}{2}A \lb( n_a \gamma_{(\xi)}{}_{bc} + n_b \gamma_{(\xi)}{}_{ca} - n_c \gamma_{(\xi)}{}_{ba}  \rb)
\ee
Then, replacing \(\nabla_b \xi_c\) using \cref{eq:D-xi} we have \cref{eq:D-chi-tensor}, and using the spinor decompositions \cref{eq:normal-defn,eq:chi-spinor,eq:Riem-spinor} we get \cref{eq:D-chi}.
\end{proof}
\end{prop}

\begin{prop}\label{prop:alpha-chi-Lie}
    For any two \(\xi_1^a, \xi_2^a \in \bms_{\bb C}\), let \(\xi^a = \Lie_{\xi_1} \xi_2^a\) be their Lie bracket. Then,
\begin{subequations}\begin{align}
    \alpha_{(\xi)} & \hateq \xi_1^b \nabla_b \alpha_{(\xi_2)} - \xi_2^b \nabla_b \alpha_{(\xi_1)} \label{eq:alpha-Lie} \\
    \chi_{(\xi)}{}_{ab} & \hateq \xi_1^c \nabla_c \chi_{(\xi_2)}{}_{ab} - \xi_2^c \nabla_c \chi_{(\xi_1)}{}_{ab} - \chi_{(\xi_1)}{}_{a}{}^c \chi_{(\xi_2)}{}_{bc} + \chi_{(\xi_2)}{}_{a}{}^c \chi_{(\xi_1)}{}_{bc} - \xi_1^c \xi_2^d R_{cdab} \label{eq:chi-Lie}
\end{align}\end{subequations}
\begin{proof}
Taking the divergence of \(\xi^a = \Lie_{\xi_1} \xi_2^a = \xi_1^b \nabla_b \xi_2^a - \xi_2^b \nabla_b \xi_1^a\) and commuting the derivatives, we see that the resulting Riemann tensor terms cancel out. The remainder of the expression directly yields \cref{eq:alpha-Lie}.

Similarly, we have (where \(1 \lra 2\) indicates the preceding terms with the labels \(1\) and \(2\) interchanged, and \(a \lra b\), similarly indicates exchange of the abstract indices)
\be\label{eq:chi-lie-temp}
    \chi_{(\xi)}{}_{ab} = \nabla_{[a} \xi_{b]} &= \half \nabla_a \lb[ \xi_1^c \nabla_c \xi_{2b} - (1 \lra 2) \rb] - (a \lra b) \\
    & = \half \nabla_a \xi_1^c \nabla_c \xi_{2b} + \half \xi_1^c \nabla_a \nabla_c \xi_{2b} - (1 \lra 2) - (a \lra b)
\ee

Using \cref{eq:D-xi} to replace \(\nabla_a \xi_c\) in the first term in \cref{eq:chi-lie-temp} gives
\be\label{eq:t1}
    \half \nabla_a \xi_1^c \nabla_c \xi_{2b} - (1 \lra 2) - (a \lra b) \hateq - \chi_{(\xi_1)}{}_{a}{}^c \chi_{(\xi_2)}{}_{bc} + \chi_{(\xi_2)}{}_{a}{}^c \chi_{(\xi_1)}{}_{1bc}
\ee
The second term in \cref{eq:chi-lie-temp} can we written as
\be
    \xi_1^c \nabla_a \nabla_c \xi_{2b} & = \xi_1^c \nabla_c \nabla_a \xi_{2b} - \xi_1^c R_{acbd} \xi_2^d
\ee
so that
\be\label{eq:t2}
    \half \xi_1^c \nabla_a \nabla_c \xi_{2b} - (1\lra 2) - (a \lra b) = \xi_1^c \nabla_c \chi_{(\xi_2)}{}_{ab} - \xi_2^c \nabla_c \chi_{(\xi_1)}{}_{ab} - \xi_1^c \xi_2^d R_{cdab}
\ee
Combining \cref{eq:chi-lie-temp,eq:t1,eq:t2} we get the desired result \cref{eq:chi-Lie}.
\end{proof}
\end{prop}

\section{Computation of the \(\mf K\)-brackets projected to \(\bb S^2\)}
\label{sec:NS-comp}

In this appendix we collect the computations showing that the projected brackets \cref{eq:proj-brackets} of the BMS superalgebra \(\mf K\) are the same as the brackets of the global NS supersymmetries on a \(2\)-sphere, as claimed in \cref{sec:NS}.

We recall the following properties of the spin-weighted spherical harmonics \cite{PR1,Stewart}. A spin-weighted spherical harmonic \(Y^s_{\ell,m}\) is a smooth function on \(\bb S^2\) satisfying
\be\label{eq:swsh-eth}
    \eth Y_{\ell,m}^s = - \sqrt{\frac{(\ell-s)(\ell+s+1)}{2}} Y_{\ell,m}^{s+1} \eqsp \eth' Y_{\ell,m}^s = \sqrt{\frac{(\ell+s)(\ell-s+1)}{2}} Y_{\ell,m}^{s-1}\,.
\ee
The harmonic \(Y^s_{\ell,m}\) is non-zero if and only if \(\abs{s} \leq \ell\) and \(\abs{m} \leq \ell\). For a given spin \(s\) the harmonics \(Y^s_{\ell,m}\) form a complete basis for the space of smooth functions with spin \(s\). An explicit expression for these harmonics in terms of the stereographic coordinates on \(\bb S^2\) can be found in \cite{Stewart}. Further, under complex conjugation we have
\be\label{eq:swsh-cc}
     \bar{Y_{\ell,m}^s} = (-)^{m+s} Y_{\ell,-m}^{-s}
\ee
The product of two harmonics can be written as a sum over other harmonics as
\be\label{eq:swsh-prod}
    Y^{s_1}_{\ell_1,m_1} Y^{s_2}_{\ell_2,m_2} = \sum_{s,\ell,m} \lb( \tfrac{(2\ell_1+1)(2\ell_2+1)(2\ell+1)}{4\pi} \rb)^\half  
    \begin{pmatrix}
    \ell_1& \ell_2 & \ell \\[-2ex]
    m_1 & m_2& m
    \end{pmatrix}
    \begin{pmatrix}
    \ell_1& \ell_2 & \ell \\[-2ex]
    -s_1 & -s_2& -s
    \end{pmatrix}
    (-1)^{m+s} Y^{-s}_{\ell,-m}
\ee
where
\(
    \begin{pmatrix}
    \ell_1& \ell_2 & \ell \\[-2ex]
    m_1 & m_2& m
    \end{pmatrix}
\)
is the \emph{Wigner \(3j\)-symbol} which is non-zero if and only if (\S~34 of \cite{DLMF})
\be\label{eq:3j-conds}
    m_1 + m_2 + m = 0 \eqsp \abs{\ell_1 - \ell_2} \leq \ell \leq \ell_1 + \ell_2 \eqsp \ell_1 + \ell_2 + \ell \text{ is an integer} \,.
\ee
The \(3j\)-symbols can be explicitly computed using eqs.~(34.24) and (34.2.5) in \S~34.2 of \cite{DLMF}, or using the function \texttt{ThreeJSymbol} in \textsc{Mathematica} \cite{Mathematica}.

Using the above properties of the spin-weighted harmonics we can show that the brackets \cref{eq:proj-brackets} reproduce those of the global NS supersymmetries \cref{eq:NS-brackets} as claimed. The basic strategy is as follows. We will substitute the basis elements \cref{eq:NS-basis} into the brackets \cref{eq:proj-brackets}, and express the right-hand-sides in terms of products of spin-weighted spherical harmonics. These products can then be expressed in terms of sums of other spherical harmonics using \cref{eq:swsh-prod,eq:3j-conds}. In each case of interest only one term in the general sum will be non-zero which can again be reexpressed in terms of the basis elements \(L_n\) and \(G_r\). Finally, we explicitly compute the \(3j\)-symbols for each case with \(n \in \{-1,0,1\}\) and \(r \in \{-\half,\half\}\) and show that the brackets match the NS brackets \cref{eq:NS-brackets}.

Lets start substituting \(L_n\) and \(L_{n'}\) into the first bracket in \cref{eq:proj-brackets}. Using \cref{eq:NS-basis,eq:swsh-eth} we get
\be
    \lb[L_n, L_{n'} \rb] = - \lb(2^{\abs n} \tfrac{4\pi}{3} \rb)^\half \lb(2^{\abs {n'}} \tfrac{4\pi}{3} \rb)^\half \lb( Y^{-1}_{1,n} Y^{0}_{1,n'} - (n \lra n') \rb)
\ee
Now using \cref{eq:swsh-prod,eq:3j-conds} we get
\be
    Y^{-1}_{1,n} Y^{0}_{1,n'} = \sum_{\ell=1}^2 \lb( \tfrac{3 \times 3 (2\ell+1)}{4\pi} \rb)^\half  
    \begin{pmatrix}
    1& 1 & \ell \\[-2ex]
    n & n' & - (n + n')
    \end{pmatrix}
    \begin{pmatrix}
    1 & 1 & \ell \\[-2ex]
    1 & 0 & -1
    \end{pmatrix}
    (-1)^{-n-n'+1} Y^{-1}_{\ell,n+n'}
\ee
Note that the \(\ell = 0\) contribution vanishes since the harmonic on the right-hand-side has spin \(s = -1\). Further, evaluating the \(3j\)-symbols one can verify that the \(\ell=2\) contribution also vanishes once antisymmetrized over \(n\) and \(n'\). So we are only left with the \(\ell=1\) contribution which gives
\be
    \lb[ L_n , L_{n'} \rb] & = \lb(2^{\abs n} \tfrac{4\pi}{3} \rb)^\half \lb(2^{\abs {n'}} \tfrac{4\pi}{3} \rb)^\half \lb(2^{\abs {n+n'}} \tfrac{4\pi}{3} \rb)^{-\half} \\
    &\qquad \times 2 
    \lb( \tfrac{3 \times 3 \times 3}{4\pi} \rb)^\half  
    \begin{pmatrix}
    1& 1 & 1 \\[-2ex]
    n & n' & - (n + n')
    \end{pmatrix}
    \begin{pmatrix}
    1 & 1 & 1 \\[-2ex]
    1 & 0 & -1
    \end{pmatrix}
    (-1)^{-n-n'+1} L_{n+n'} \\
    & = (n - n') L_{n+n'}
\ee
where in the final step we evaluate the \(3j\)-symbols for the values \(n,n' \in \{-1,0,1\}\).

Similarly, we subsitute two basis elements \(G_r\) and \(G_{r'}\) (\cref{eq:NS-basis}) into the last bracket in \cref{eq:proj-brackets} to get
\be
    \lb[ G_r, G_{r'} \rb] = - 4 \sqrt{2} \pi~ Y^{-\half}_{\half,r} Y^{-\half}_{\half,r'} 
\ee
Again using \cref{eq:swsh-prod,eq:3j-conds} we get a harmonic of spin \(s = -1\) and only a \(\ell=1\) term. This gives
\be
    Y^{-\half}_{\half,r} Y^{-\half}_{\half,r'} = \lb( \tfrac{(2)(2) 3}{4\pi} \rb)^\half  
    \begin{pmatrix}
    \half & \half & 1 \\[-2ex]
    r & r' & -(r+r')
    \end{pmatrix}
    \begin{pmatrix}
    \half & \half & 1 \\[-2ex]
    \half & \half & -1
    \end{pmatrix}
    (-1)^{-r-r'+1} Y^{-1}_{1,r+r'}
\ee
So
\be
    \lb[ G_r, G_{r'} \rb] & = 4 \sqrt{2} \pi \lb(2^{\abs{r+r'}} \tfrac{4\pi}{3} \rb)^{-\half}
    \lb( \tfrac{(2)(2) 3}{4\pi} \rb)^\half  
    \begin{pmatrix}
    \half & \half & 1 \\[-2ex]
    r & r' & -(r+r')
    \end{pmatrix}
    \begin{pmatrix}
    \half & \half & 1 \\[-2ex]
    \half & \half & -1
    \end{pmatrix}
    (-1)^{-r-r'+1} L_{r+r'} \\
    & = 2L_{r+r'}
\ee
where to get the last line we evaluate the \(3j\)-symbols for \(r,r' \in \{-\half,\half\}\).

Lastly, we substitute an \(L_n\) and a \(G_r\) into the second bracket in \cref{eq:proj-brackets}, and use \cref{eq:swsh-eth} to obtain
\be
    \lb[L_n, G_r\rb] &= - \lb(2^{\abs n} \tfrac{4\pi}{3} \rb)^\half \lb( -i 2 \sqrt{2} \pi \rb)^\half \lb[ - \tfrac{1}{\sqrt{2}} Y^{-1}_{1,n} Y^{\half}_{\half,r} + \tfrac{1}{2} Y^{-\half}_{\half, r} Y^{0}_{1,n} \rb]
\ee
Evaluating both products using \cref{eq:swsh-prod,eq:3j-conds} gives a \(Y^{-\half}_{\ell,n+r}\) and the sum over \(\ell \in \{\half, \nfrac{3}{2}\}\). An explicit computation shows that the \(3j\)-symbols in the square brackets cancel for \(\ell = 3/2\), so we are only left with \(\ell=\half\) which gives
\be
    \lb[ L_n, G_r \rb] &= - \lb(2^{\abs n} \tfrac{4\pi}{3} \rb)^\half
    \lb( \tfrac{(3)(2)(2)}{4\pi} \rb)^\half  
    \begin{pmatrix}
    1& \half & \half \\[-2ex]
    n & r & -(n+r)
    \end{pmatrix}
    (-1)^{-n-r+\half} G_{n+r} \\
    &\qquad \times \lb[
    - \tfrac{1}{\sqrt{2}} \begin{pmatrix}
    1 & \half & \half \\[-2ex]
    1 & -\half & -\half
    \end{pmatrix}
    + \tfrac{1}{2} \begin{pmatrix}
    1 & \half & \half \\[-2ex]
    0 & \half & -\half
    \end{pmatrix} \rb] \\
    &= (n/2 - r) G_{n+r} 
\ee
where the last line follows from evaluating the \(3j\)-symbols for \(n \in \{-1,0,1\}\) and \(r \in \{-\half,\half\}\).



\bibliographystyle{JHEP}
\bibliography{twistors-bms}
\end{document}